\documentstyle[11pt,aaspp4]{article}
\eqsecnum
\def\beq{\begin{equation}}
\def\eeq{\end{equation}}
\def\ref{\reference}
\def\simge{\mathrel{%
   \rlap{\raise 0.511ex \hbox{$>$}}{\lower 0.511ex \hbox{$\sim$}}}}
\def\simle{\mathrel{
   \rlap{\raise 0.511ex \hbox{$<$}}{\lower 0.511ex \hbox{$\sim$}}}}

\begin{document}
\title{Spectral Transition and Torque Reversal in X-ray Pulsar 4U 1626-67}
\author{Insu Yi\altaffilmark{1,2,3} and Ethan T. Vishniac\altaffilmark{4}}

\altaffilmark{1}{Center for High Energy Astrophysics and Isotope Studies,
Research Institute for Basic Sciences}

\altaffilmark{2}{Dept. of Physics, Ewha University, Seoul, Korea;
yi@astro.ewha.ac.kr}

\altaffilmark{3}{Korea Institute for Advanced Study, Seoul, Korea}

\altaffilmark{4}{Dept. of Physics and Astronomy, Johns Hopkins University, 
Baltimore, MD 21218; ethan@pha.jhu.edu}

\begin{abstract}

The accretion-powered, X-ray pulsar 4U 1626-67 has recently shown an abrupt
torque reversal accompanied by a dramatic spectral transition and a relatively
small luminosity change. The time-averaged X-ray spectrum during spin-down is 
considerably harder than during spin-up. The observed torque reversal can be 
explained by an accretion flow transition triggered by a gradual change 
in the mass accretion rate. The sudden transition to spin-down is caused by 
a change in the accretion flow rotation from Keplerian to sub-Keplerian. 
4U 1626-67 is estimated to be near spin equilibrium with a mass accretion 
rate ${\dot M}\sim 2\times 10^{16}$g/s, ${\dot M}$ decreasing at a rate 
$\sim -6\times 10^{14}$ g/s/yr, and a polar surface magnetic field of 
$\sim 2b_p^{-1/2}\times 10^{12}$G where $b_p$ is the magnetic pitch. 
During spin-up, the Keplerian flow remains geometrically thin and cool. 
During spin-down, the sub-Keplerian flow becomes geometrically thick and hot. 
Soft photons from near the stellar surface are Compton up-scattered by the 
hot accretion flow during spin-down while during spin-up such scattering 
is unlikely due to the small scale-height and low temperature of the flow. 
This mechanism accounts for the observed spectral hardening and small 
luminosity change. The scattering occurs in a hot radially falling column 
of material with a scattering depth $\sim 0.3$ and a temperature $\sim 10^9K$. 
The X-ray luminosity at energies $>5$keV could be a poor indicator of the mass 
accretion rate. We briefly discuss the possible application of this mechanism 
to GX 1+4, although there are indications that this system is significantly 
different from other torque-reversal systems.

\end{abstract}


\vfill\eject

\section{Introduction}

The continuous monitoring of the accretion-powered X-ray pulsar 4U 1626-67 by
BATSE (e.g. Chakrabarty et al. 1997a) has revealed that the pulsar system has 
recently undergone a puzzling, abrupt torque-reversal from spin-up to spin-down.
Similar phenomena have also been detected in other pulsar systems 
(e.g. Bildsten et al. 1997). The X-ray luminosity in the 1-20 keV band changes
little around the torque reversal (Vaughan \& Kitamoto 1997), which indicates 
that the reversal event is not likely to be caused by a large, discontinuous 
change of mass accretion rate (Yi et al. 1997, Yi \& Wheeler 1998).
This phenomenon is difficult to explain within the context of a Ghosh-Lamb
magnetized disk model (Ghosh \& Lamb 1979) in which the torque and mass
accretion rate would be expected to change together. 

Nelson et al. (1997) proposed that it is due to a sudden change in the sense 
of accretion disk rotation with a nearly constant mass accretion rate. 
This possibility seems suspect given that the accretion flow is apparently 
stable up until the reversal event and then has to suddenly reverse its 
rotational direction. The formation of a retrograde disk is more likely in 
wind-fed systems in which small fluctuations due to fluctuating angular 
momentum are apparent observationally (e.g. Nagase 1989).
Van Kerkwijk et al. (1998) have suggested an intriguing possibility in which
the inner disk is warped by an irradiation instability to such an extent 
that the tilt angle becomes larger than 90 degrees and the accretion flow's 
rotation becomes retrograde. In this explanation, there exist several 
outstanding issues such as (i) the uncertain flip time scale, 
(ii) the return mechanism from the high tilt back to low tilt angle, 
(iii) the X-ray irradiation efficiency in strongly magnetized accretion 
(e.g. Yi \& Vishniac 1994), and (iv) the severe X-ray obscuration by the 
tilted disk material (van Kerkwijk et al. 1998).

The observed X-ray luminosity appears to decrease slightly from spin-up to 
spin-down. In the 1-20 keV GINGA band, the X-ray luminosity decreases by about 
$\sim 20$\%. This flux decrease seems to be more significant at lower energies 
within this band (Vaughan \& Kitamoto 1997). During spin-up, the 
phase-averaged spectrum can be modeled by a blackbody with a temperature 
$\sim 0.6$ keV (Angelini et al. 1995) together with a power-law 
component with a spectral index $\sim 1$ (Kii et al. 1986). 
Vaughan and Kitamoto (1997) discuss simple power-law fits to the time-averaged 
GINGA ASM count data. The strong energy dependence of the pulse profile 
indicates anisotropic radiative 
transfer within a magnetically channeled accretion column with a strong 
magnetic field $>6\times 10^{12} G$ (e.g. Kii et al. 1986). Recently, BeppoSAX 
(0.1-10 keV) detected X-ray emission during the spin-down phase 
(Orlandini et al. 1997, Owens et al. 1997). During this phase
the spectrum is well fit by an absorbed power-law of index $\sim 0.6$ 
and a blackbody of temperature $\sim 0.3$ keV. The absorption column depth 
is $\sim 10^{21} cm^{-2}$ (Owens et al. 1997). Vaughan \& Kitamoto (1997) 
suggest much harder spectra during spin-down with a power law index 
$\sim 0.41$, although $\sim 0.6$ is within their allowed spectral index range. 
GINGA spin-up spectra are described by the power-law index of $\sim 1.48$ which
is somewhat steeper than the index seen in mid- to late 1980s. This could well 
be the result of the wider energy band. A self-consistent explanation for the 
observed spectral transition compatible with the torque reversal, is required. 
Absorption has been suggested as a possible explanation. 
The required absorption column 
density is however as high as $N_H\approx 10^{23}-10^{24} cm^{-2}$ 
(Vaughan \& Kitamoto 1997), which is inconsistent with the relatively low
accretion rate. On the other hand, in the model of van Kerkwijk et al. (1998), 
as the disk tilt angle increases, absorption by the disk is expected 
to be too severe to account for the observed X-ray spectra.

In this paper, we propose that the spectral change is directly related to 
the formation of a geometrically thick hot accretion flow during the 
spin-down, which Compton up-scatters the soft photons emitted near the 
stellar surface (Yi et al. 1997). Such a hot flow is available only during 
spin-down, where the spin-down itself is the result of the sub-Keplerian 
rotation of the hot accretion flow (e.g. Narayan \& Yi 1995). 

\section{Torque Reversal, Mass Accretion Rate, and X-ray Luminosity}

In a conventional disk-magnetosphere interaction model with a Keplerian
accretion flow with the mass accretion rate ${\dot M}$, 
the torque exerted on the pulsar is
\beq
N=(7N_0/6)\left[1-(8/7)(R_0/R_c)^{3/2}\right]\left[1-(R_0/R_c)^{3/2}\right]^{-1}
\eeq
where $R_c=(GM_*P_*^2/4\pi^2)^{1/3}$ is the Keplerian corotation radius for
a pulsar of mass $M_*=1.4M_{\sun}$ and spin period $P_*$ and
$N_0={\dot M}(GM_*R_0)^{1/2}$ is the material torque at the disk disruption
radius $R_0$, which is determined by
\beq
(R_0/R_c)^{7/2}\left[1-(R_0/R_c)^{3/2}\right]^{-1}=2(2\pi)^{7/3}b_p
B_*^2R_*^5/(GM_*)^{2/3}P_*^{7/3}L_x=\delta.
\eeq
$b_p$ is a parameter of order unity which sets the pitch of the magnetic 
field (Wang 1995, Yi et al. 1997, and references therein). 
$B_*$ is the polar surface magnetic field strength of 
the pulsar and $L_x\approx GM_*{\dot M}/R_*$ is the X-ray luminosity from the 
surface of the accreting pulsar with radius $R_*$ and accretion rate 
${\dot M}$. 
In this picture, as ${\dot M}$ varies, the pulsar spin period changes 
according to ${\dot P}_*/P_*^2=-N/2\pi I_*$ where $I_*=10^{45} g cm^2$ is the 
neutron star moment of inertia. If ${\dot M}$ variation is smooth, 
the corresponding torque change is also smooth (e.g. Yi et al. 1997). 
The torque decreases in response 
to the decreasing ${\dot M}$. In this case, the torque should
show a wide range of positive and negative values and pass through $N=0$.
This is apparently not observed in 4U 1626-67 and other similar systems
(Chakrabarty 1997a, Nelson et al. 1997). 
The observed spin-down to spin-up transition is too abrupt to be reproduced 
by gradual ${\dot M}$ change in the Keplerian models. The observed 
event could be reproduced only if the mass accretion rate jumps by a factor
$\sim 4$ almost discontinuously (Yi et al. 1997), which is highly unlikely
given the fact that the observed luminosity decreases only by about 
$\sim 20$\% (Vaughan \& Kitamoto 1997). 

The model proposed by Yi, Wheeler, \& Vishniac (1997) suggests that the torque 
reversal is due to the transition of the accretion flow from (to) Keplerian 
rotation to (from) sub-Keplerian rotation. The transition occurs when 
${\dot M}$ crosses the critical rate ${\dot M}_c$ which lies somewhere
in a range $\sim 10^{16}-10^{17}$ g/s. 
After the transition, the rotation of the accretion flow becomes sub-Keplerian 
due to the large internal pressure of the hot accretion flow.
For sub-Keplerian rotation, $\Omega=A\Omega_K<\Omega_K$ 
(i.e. $A<1$ and $\Omega_K=(GM_*/R^3)^{1/2}$ is the Keplerian rotation), 
the new corotation radius
$R_c^{\prime}=A^{2/3}R_c$ and the new inner edge $R_0^{\prime}$ is determined 
by
\beq
(R_0^{\prime}/R_c^{\prime})^{7/2}\left[1-(R_0^{\prime}/R_c^{\prime})^{3/2}
\right]^{-1}=2(2\pi)^{7/3}b_pA^{-7/3}B_*^2R_*^5/(GM_*)^{2/3}P_*^{7/3}L_x
=\delta^{\prime}.
\eeq
The new torque on the star after the transition to the sub-Keplerian rotation 
is
\beq
N^{\prime}=(7N_0^{\prime}/6)
\left[1-(8/7)(R_0^{\prime}/R_c^{\prime})^{3/2}\right]
\left[1-(R_0^{\prime}/R_c^{\prime})^{3/2}\right]^{-1},
\eeq
where $N_0^{\prime}=A{\dot M}(GM_*R_o^{\prime})^{1/2}$.
The reversal is simply a consequence of the change in $A$ from 1 to $<1$ 
(Yi et al. 1997) as ${\dot M}$ gradually crosses ${\dot M}_c$. 
It has been shown that the observed torque reversals occur near 
spin-equilibrium (i.e. $N=0$) before and after transition (Yi et al. 1997) 
satisfying
$b_p^{1/2}B_*\approx 7\times 10^{11}\delta^{1/2}(L_x/10^{36}erg/s)^{1/2}
(P_*/10s)^{7/6}~G$
where $2A^{7/6}<\delta^{1/2}<4A^{7/6}$ is likely to be of order unity
(Yi \& Wheeler 1998).

The observed spin frequency evolution is easily accounted for by
${\dot M}_c=2.4\times 10^{16} g/s$, $A=0.462$ at torque reversal with 
$B_*=5.4(b_p/10)^{-1/2}\times 10^{11}G$, which are consistent with
the above criterion (Yi \& Wheeler 1998). 
The model requires only a small change of 
${\dot M}$ at a rate $d{\dot M}/dt\sim -6\times 10^{14} g/s/yr$, about a 
factor of $5$ less than the required $d{\dot M}/dt$ for the Keplerian model. 
The observed small flux change 
$\sim 20$\% around the torque reversal is consistent with the present model.
This model is yet to explain the negative second derivative of the spin
frequency before and after the reversal (Chakrabarty et al. 1997a).
The truncation radius changes from $R_o\sim 5.5\times 10^8cm$ just before
reversal to $R_o^{\prime}\sim 3.7\times 10^8cm$ right after reversal. 
We look for a mechanism for the spectral transition 
with small changes in ${\dot M}$ within the model of Yi et al. (1997).
The small change in the integrated X-ray luminosity suggests a
comparably small change in the mass accretion rate. In the magnetized 
accretion picture, if the accretion flow is truncated far away from the 
neutron star, most of the X-ray emission should originate in the column 
accretion flow near the neutron star. In this case, the X-ray luminosity is
likely to reflect the mass accretion rate change. Since the observed spectra 
show a strong thermal emission component, it is plausible that the X-ray 
emission comes from the thermalized material near the accreting magnetic 
poles (e.g. Kaminker et al. 1976, Frank et al. 1992).

When the accretion flow is Keplerian, 
the accretion flow is dominated by cooling and the accreted plasma remains
cold until it undergoes shock heating and thermalization near the 
magnetic poles.  In this case, the accretion column is not
hot enough for significant Compton scattering. The cold accretion
column is more likely to act as absorbing gas than a Comptonizing
gas. The geometric thickness of the accretion column flow is
much smaller than the stellar radius, so that only a small fraction
of the radiation from the star will interact with the accreted material.
After transition, the accretion flow is hot and its internal 
pressure becomes large enough to support a large scale height. The accretion 
flow extending from $R_o^{\prime}$ to $R_*$ is geometrically thick and hot 
enough to Comptonize the thermal stellar radiation.

\section{Scattering in Hot Accretion Column during Spin-down}

The sub-Keplerian flow is advection-dominated (e.g. Narayan \& Yi 1995), 
which implies that the bulk of the viscously dissipated energy is stored 
within the accretion flow. Assuming that advection is the dominant
channel for the viscously dissipated energy, the temperature of the accretion 
flow is $T_{acc}\approx 1.2\times 10^9 K$ and the electron scattering
depth along the vertical direction is $\tau_{es, acc}\approx 2.8\times 10^{-2}$ 
at $R=R_o^{\prime}\approx 4\times 10^8cm$.  Here we have assumed an
equipartition magnetic field and a viscosity parameter $\alpha=0.3$ 
(Narayan \& Yi 1995). This implies that the advection-dominated accretion 
flow itself has a very small scattering depth.
However, the accretion flow inside the radius $R=R_o^{\prime}$ is 
channeled by the magnetic fields lines, and is a very promising site for 
scattering due to geometric focusing. In the conventional column accretion
model (e.g. Frank et al. 1992, Yi \& Vishniac 1994), if the dipole magnetic 
field axis is misaligned with respect to rotation axis of the star 
(which is assumed to be aligned with the accretion flow's rotational axis) 
by an angle $\pi/2-\beta_1$, then the angle between the magnetic axis and the 
magnetic field lines connected to the disk plane at $R_o^{\prime}$,
$\beta_2$, is given by $\sin^2\beta_1=\sin^2\beta_2\times (R_*/R_o^{\prime})$. 
It can also be shown that the fraction of the polar surface area threaded by
magnetic field lines connected to the accretion flow at $R\ge R_o^{\prime}$ is
$s\approx R_*\sin^2\beta_2/4R_o^{\prime}\approx R_*/2R_o^{\prime}$ after 
angle-averaging over $\beta_2$. This implies that unless $\beta_2$ is very 
small the accreting fraction of the stellar surface area $s$ is likely to be 
of order $\sim R_*/R_o^{\prime} \sim 10^{-2}-10^{-3}$.
The accretion is assumed to continue to the poles without mass loss and then
the soft X-ray radiation from the polar regions is uniquely determined 
by the mass accretion rate derived from the torque reversal (Yi \& Wheeler
1998). 

The detailed emission process near the magnetic poles is complicated 
(e.g. Frank et al. 1992 and references therein).
The existence of a thermal component at low X-ray energies
implies that a significant fraction of the accretion energy is thermalized
near the surface of the neutron star. If the entire accretion energy is
thermalized at the stellar surface (cf. Kaminker et al. 1976), 
the anticipated blackbody temperature is $kT_*\sim 5s^{-1/4} keV$. 
Since the accretion rate change is small near the torque reversal, the observed
spectral transition is unlikely to be caused by the sudden change in physical 
conditions. If there is no scattering effect from the infalling
column material, the underlying emission process is likely to remain
unchanged around the torque reversal. We therefore assume that the intrinsic
spectrum (before scattering and absorption) remains unchanged.

The column accretion flow follows the bent magnetic field lines. Since a 
detailed geometry is hard to specify, we assume that the solid angle subtended 
by the accretion column at the accretion poles is constant at all radii along 
the field lines. The radial bulk infall velocity can be parametrized as a 
fraction of the free-fall velocity as $v_R=-x(2GM_*/R)^{1/2}$ and we get 
\beq
\tau_{es}\approx {\sigma_T\over 4\pi\mu m_p}{1\over s^{1/2} x}
{{\dot M}\over (2GM_*R)^{1/2}}
\eeq
where $\mu\sim 1$ is the mean molecular weight.
We find that $\tau_{es}\sim 0.4\mu (R/10^6cm)^{-1/2}$ with
$s\sim 10^{-2}$, $x\sim 1$, ${\dot M}\sim 2.4\times 10^{16} g/s$
where the mass accretion rate is based on the torque reversal fitting
(e.g. Yi \& Wheeler 1998). From the inner disk radius 
$R=R_o^{\prime}\sim 4\times 10^8 cm$ to the base of the accretion column 
near the stellar surface $R\sim 10^6 cm$, the scattering depth is 
likely to range from $\tau_{es}\sim 2\times 10^{-2}\mu^{-1}$ to 
$\sim 0.1\mu^{-1}$ at $R\sim 10^7 cm$ to $\sim 0.4\mu^{-1}$ at $R\sim 10^6cm$. 
Evidently the accretion column can have a dramatic effect on the outgoing soft
radiation through scattering.

The angular momentum has to be continuously transported along the accretion 
column (corresponding to the torque action between the star and accretion flow).
We assume the radial infall is nearly adiabatic and the shear dissipation is
decoupled from the bulk radial motion.
We take the dissipation rate per unit volume within the accretion column as
$q^{+}=\nu \rho R^2 (d\Omega/dR)^2$
where $\nu=c_s H$ is the kinematic viscosity and $H\sim 2\sqrt{s}R$ is the
cross-sectional thickness of the accretion column. We have assumed that the
internal random energies, thermal and turbulent, are comparable and the
bulk radial motion is supersonic. We then obtain
$q^{+}\sim 18G\sqrt{s} M_* c_s\rho/4r$
or the integrated dissipation rate per unit length of the column
$Q^{+}\sim q^+\times \pi D^2\sim 9c_s |v_R| {\dot M} D^2/8 R^2$.
The cooling of the column material occurs via Compton up-scattering
of the soft photons from the thermalized radiation near the stellar 
surface (e.g. Narayan \& Yi 1995).
Assuming that the Compton interaction occurs at a distance sufficiently
far away from the stellar surface, we treat the thermal radiation as
originating from a point source. Then the integrated cooling rate across
the column is estimated as
$Q^{-}\sim 4kGM_*{\dot M}D\tau_{es}T/2m_e c^2 R_*R^2$.
The energy balance $Q^{+}=Q^{-}$ gives
$(kT)^{1/2}\tau_{es}\sim 9 m_e c^2R_*D/16(\mu m_p)^{1/2}\sqrt{s} |v_R| R^2$
which shows the expected inverse correlation between the Compton depth
and the column temperature. In general, $Q^+=Q^-$ is not guaranteed but
as $\tau_{es}\rightarrow 1$ near the stellar surface (e.g. Yi \& Vishniac
1994), a significant Compton cooling could lead to $Q^+=Q^-$ (cf. Narayan
\& Yi 1995). In this simple picture, the column
temperature remains nearly independent of $R$ until the column material 
becomes optically thick and the radiation is thermalized.
For parameters $s\sim 10^{-2}$, and ${\dot M}\sim 10^{16} g/s$, we expect
$kT\sim 60 keV$. This implies that the resulting Comptonized radiation 
is likely to show the signature of a hot plasma with this temperature.
The sub-Keplerian accretion flow at $R>R_o^{\prime}$ 
is advection-dominated (e.g.  Narayan \& Yi 1995). 
At radii $R>R_o^{\prime}$, the temperature difference between
electrons and ions is likely to be small. After the accretion flow enters
into the magnetic column, the accreted material is rapidly compressed, which
justifies the single temperature treatment of the energy balance between
viscous heating and Compton cooling.
The Comptonized radiation is calculated according to $E_f=\eta E_i$
where $E_f$ the photon energy after scattering and $E_i$ is the
incident soft energy. $\eta$ is the energy boost parameter originally
derived by Dermer et al. (1991), i.e. $\eta=1+\eta_1-\eta_2(E_i/kT)^{\eta_3}$
where $\eta_1$, $\eta_2$, and $\eta_3$ are defined in Narayan \& Yi (1995). 
These functions are completely specified by $\tau_{es}$ and $T$ for
a given $E_i$. Therefore, given the incident photon spectrum $N_i(E_i)$
the resulting Comptonized spectrum is completely specified.
The radial infall speed in the scattering region and its effects on
the scattering could be significant, which is beyond the scope of this work. 

Although the soft radiation is likely to be the thermalized radiation
from the stellar surface, the details of the emission remains unclear.
We simply assume that the soft photon spectrum remains unchanged near the 
torque reversal and that the unscattered soft photon spectrum during spin-down 
is similar to that of the spin-up period. Since during spin-up scattering is
negligible, this assumption is quite plausible. We further assume that the 
luminosity of the source decreases according to the mass accretion rate decline.
We expect the soft photon luminosity to decrease by a few $\times 10$\% while 
the spectral shape remains unchanged. We also assume that all outgoing 
radiation goes through the Comptonizing accretion column.
Figure 1 shows the result of the Compton scattering from a column
material with $T_e=10^9K$ and $\tau_{es}\sim 0.3$ which are close to the
values we have derived above. These parameters adequately reproduce the
qualitative features of the observed spectral transition. The shown spectral
transition requires that the soft photon flux decrease by $\sim 40$\% 
during spin-down, which is not far from the change required for the torque 
reversal (Yi et al. 1997, Yi \& Wheeler 1998). 
Despite the uncertainties in accretion column geometry, 
the expected spectral change is remarkably close to the observed change. 
This strongly indicates that the observed spectral transition could be
due to formation of the hot accretion flow during spin-down. More detailed, 
quantitative answers would require exact scattering geometric information 
and physics of thermal radiation emission near the stellar surface. 

\section{Discussion}

We have shown that Compton-scattering during spin-down could account
for the sudden spectral transition correlated with the torque reversal
seen in 4U 1626-67. The small luminosity change supports the possibility
that the mass accretion rate varies little as the present model indicates.
The expected beat-type QPOs would be hard to detect during spin-down 
(Yi \& Wheeler 1998). Kommers et al. (1998) reported the detection of QPOs 
at a frequency of 0.048Hz using RXTE, but they found that the observed QPOs 
are not likely to be attributable to a magnetospheric beat.

Although the spectral transition appears robust, the time-averaging of
the X-ray spectra over $\sim 30$ days (Vaughan \& Kitamoto 1997) could be 
questionable due to some changes in accretion flows near the torque reversal
on short time scales.
If the accretion flow transition occurs gradually over a period of days 
(e.g. Yi et al. 1997), the early spin-down spectra could be substantially 
softer than late spin-down spectra. If such a difference is not found, 
the accretion flow transition should occur on a time scale $\simle day$. 
Such a short transition time scale is still plausible within the proposed model 
(Yi et al. 1997) although the details of the transition process have not 
been addressed (Yi \& Wheeler 1998). On long time scales, Owens et al.
(1997) found that the X-ray flux detected by BeppoSax is about a factor
2 lower than the ASCA flux measured three years earlier (Angelini et al.
1995). This suggests that the mass accretion rate does decrease on long time
scales although the accretion rate decrease on short time scales near the
reversal is not large enough to drive the torque reversal and spectral 
transition.

Orlandini et al. (1998) reported an absorption feature at $\sim 37$keV, which
they claim to be the cyclotron absorption line. The estimated surface
field strength is $\sim 3\times 10^{12}$G. This field strength is substantially
higher than our estimate $\sim 5(b_p/10)^{-1/2}\times 10^{11}$G. 
Such a discrepancy is not unexpected given the uncertainties in the distance 
estimate. If the magnetic pitch $b_p$ is as small as $\sim 1$,
the Orlandini et al. (1998) estimate is largely consistent with our estimate.

GX 1+4 has shown a similar torque reversal and an unexplained torque-flux 
correlation, which is exactly opposite to that predicted in the conventional
model (Chakrabarty et al. 1997b). A significant spectral hardening similar to
that seen in 4U 1626-67 could account for the correlation. So far such a 
transition has not been reported in GX 1+4. The prograde-retrograde transition 
(Nelson et al. 1997) could provide an explanation (van Kerkwijk 1998) 
despite several outstanding difficulties.

\acknowledgments
We acknowlege the support of the SUAM Foundation and KRF
grant 1998-001-D00365 (IY), NASA grant NAG5-2773 and NSF grant 
AST-9318185 (ETV). ETV is grateful for the hospitality of MIT and the CfA 
during the completion of this work. IY thanks Josh Grindlay and Craig 
Wheeler for discussions on LMXBs and pulsars.

\vfill\eject


Figure 1: Expected X-ray photon spectra before and after torque reversal. 
The scattering region used in the theoretical post-reversal spectrum has
$\tau_{es}=0.3$ and $T_e=10^9K$. The un-scattered soft radiation changes
luminosity by 40\% while its spectral shape remains unchanged (see text).

\vfill\eject
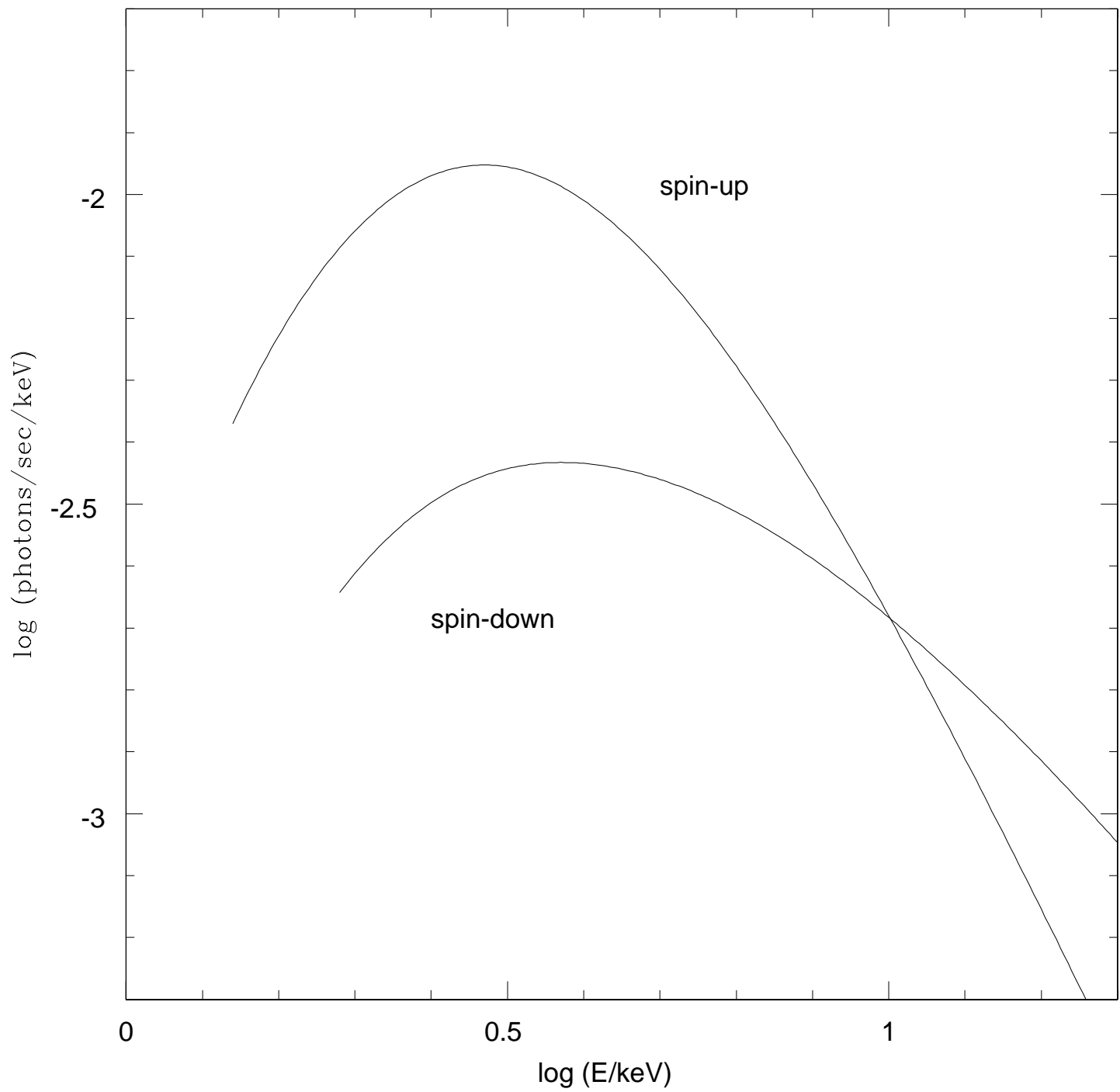
\end{document}